\documentclass[12pt]{article}

\usepackage{graphicx,amsmath,amssymb,url}
\usepackage{float}

\newlength{\dinwidth}
\newlength{\dinmargin}
\def\lapproxeq{\lower .7ex\hbox{$\;\stackrel{\textstyle
<}{\sim}\;$}}
\def\gapproxeq{\lower .7ex\hbox{$\;\stackrel{\textstyle
>}{\sim}\;$}}
\def\gtrsim{\lower .7ex\hbox{$\;\stackrel{\textstyle
>}{\sim}\;$}}
\def\lesim{\lower .7ex\hbox{$\;\stackrel{\textstyle
<}{\sim}\;$}}
\def\be{\begin{equation}}
\def\ee{\end{equation}}

\begin{document}

\titlepage

\begin{flushright}

IPPP/16/95\\

\today\\

\end{flushright}

\vspace*{0cm}

\begin{center}

{\Large \bf Updates of the MMHT2014 PDFs\footnote{Talk presented by A.D.Martin at Diffraction 2016, Acireale, Sicily, Sept. 2-8, 2016, to be published in AIP conference Proceedings.}}

\vspace*{1cm} 

{\sc L.A. Harland-Lang}$^{a}$, {\sc A.D. Martin}$^b$ and  {\sc
  R.S. Thorne}$^{a}$ \\

\vspace*{0.5cm}
$^a$ {\em Department of Physics and Astronomy, University College London, WC1E 6BT, UK}\\ 
$^b$ {\em Institute for Particle Physics
  Phenomenology, Durham University, Durham DH1 3LE, UK}\\

 \end{center}

\begin{abstract}
 We briefly discuss some of the developments since the publication of the MMHT14 parton distributions. In particular we explore the impact of recent LHC data for $W^\pm,Z$ and $t\bar{t}$ production, and perform a preliminary new analysis including these data. In this re-fit (which we tentatively call `MMHT16') there are few changes of significance in the central values of the PDFs, but some data reduce the uncertainties, mainly in the strange and valence quark distributions. We find that an extended $\bar{d}-\bar{u}$ parametrization only leads to minor changes, with the difference going to zero as $x \to 0$. We comment on the determination of the photon PDF.
\end{abstract}

\section{Introduction}

The MMHT14 PDFs \cite{MMHT14} are the successor to the MSTW08 parton distributions \cite{MSTW}. Briefly the improvements then made were (i) the parametrization of the input distributions were in terms of Chebyshev polynomials, giving more stability to the parameter values, (ii) the deuteron corrections were parametrized and the values of the parameters were determined by the fit, (iii) there was a multiplicative treatment of the errors, (iv) the nuclear corrections were updated, (v) the optimal General Mass-Variable Flavour Number Scheme was used \cite{Thorne}  and (vi) the experimental value $B_\mu =0.092\pm 10$\%$ $ of the $D \to \mu$ branching ratio was input in the fit (whereas MSTW08 used the fixed 0.099 NuTeV value).
Improvements (i) and (ii) had already been implemented in an intermediary publication \cite{MMSTWW}. The differences between the MSTW08 and MMHT14 PDFs are small -- an exception is $(u_V-d_V)$ at low $x$ which is constrained by more precise $W^\pm$ charge asymmetry data. Improvement (vi) found $B_\mu=(0.085-0.091)\pm 15$\%$ $, which results in the uncertainty and the value of $(s+\bar{s})$ distribution being increased.

The new data fitted in MMHT14 (as compared to the MSTW08 global fit) were the HERA I combined data, the updated Tevatron $W^{\pm},~Z$ data and the LHC data available then. Below we shall discuss a preliminary new global fit (`MMHT16') which includes the HERA I+II combined data and recent LHC data, particularly those on $W^\pm$ and $Z$ production.

\section{Updates of MMHT2014 PDFs already published}

There are three published extensions of the MMHT14 analysis.  The first is a study of the role of $\alpha_S$ in the analysis \cite{MMHT1}. PDF sets in an extended range of fixed values of $\alpha_S$ (about the best-fit value) were made available. This allows the error due to $\alpha_S$ to be added in quadrature in any predictions made using MMHT14 PDFs. The best-fit values in the NNLO and NLO global analyses were $\alpha_S(M_Z^2)=0.1172\pm 0.0013$ and $0.1201\pm 0.0015$ respectively.

The second paper \cite{MMHT2} investigated the variation in the MMHT14 PDFs when the heavy quark masses $m_c$ and $m_b$ were varied from their default values of 1.4 and 4.75 GeV respectively. The predictions of standard processes at the LHC show the effects of varying $m_c$ are small but not insignificant, whereas varying $m_b$ is largely insignificant, except for the $b$ quark PDF itself.

The third paper \cite{MMHT3} examined the impact of the final HERA combination of inclusive cross section data \cite{HERA}.
Already the MMHT14 predictions describe these data very well, particularly at NNLO. Consequently the inclusion of these data has little impact on the central values of the PDFs, though they do reduce the uncertainty of the PDFs, mainly for the gluon. The improvement in the uncertainty is more noticeable in the predictions of the benchmark LHC cross sections, with the uncertainty in Higgs production being 10$\%$ smaller. This paper also investigates the effects of varying $Q^2_{\rm min}$. It was also noted that the low $x$, low $Q^2$ HERA data 
can be accommodated by a power correction to $F_L$, namely $F_L\to F_L(1+a/Q^2)$ with $a\sim 4.3$ GeV$^2$, see also \cite{Mandy}.  No similar modification to $F_2$ was found to be preferred.

\section{Impact of new LHC data}
 \begin{table}[h]
\begin{center}
\begin{tabular}{|l|c|c|c|c|}
\hline
 & & MMHT14 &   `MMHT16'  \\
          & no. points    & $\chi^2$(pred.)  & $\chi^2$(fit)  \\
\hline
$\sigma(t\bar{t})$ Tevatron+CMS+ATLAS      &    18   & 14.7 &  15.5  \\
LHCb 7 TeV $W^\pm, Z$ \cite{LHCbW7latest}  & 33 & 37.1 & 36.7 \\
LHCb 8 TeV $W^\pm,~Z$ \cite{LHCbWZ8} & 34 & 76.1 & 67.2 \\
LHCb 8 TeV $Z\to e^+e^-$ \cite{LHCbWZ8}  & 17 & 30.0 & 27.8 \\
CMS 8 TeV $W^\pm$ \cite{CMSW} & 22 &57.6 & 29.4 \\
CMS 7 TeV $W+c$ \cite{CMSWc} & 10 & 8.7 & 8.0 \\
D0 $e$ charge asymmetry \cite{D0e} & 13 & 27.3 & 22.9 \\
\hline                                                        
total      &      3405          &   3768.0  & 3739.3  \\
\hline
\end{tabular}
\caption{\sf The NNLO description of new LHC data}
\vspace{-0.4cm}
\label{tab:a}
\end{center}
\end{table}

 In Table~\ref{tab:a} we show the predicted values of $\chi^2$ for LHC data not included in the NNLO MMHT2014 global analysis, together with the $\chi^2$ values when these data are included in a new global analysis - the preliminary MMHT2016 fit. The MMHT14 predictions are remarkably good, as is to be expected by comparing the values of $\chi^2$(MMHT16) with $\chi^2$(MMHT14). Examples are shown in Fig.~\ref{fig:LHCbWZ}. The only exception are the CMS $W^\pm$ data points at small rapidity, see Fig.~\ref{fig:CMSW}.

 \begin{figure} [H]
 \begin{center}
 \label{fig:LHCbWZ}
 \includegraphics[clip=true,trim=.0cm 0.0cm 0.0cm 0.8cm,height=10.4cm]{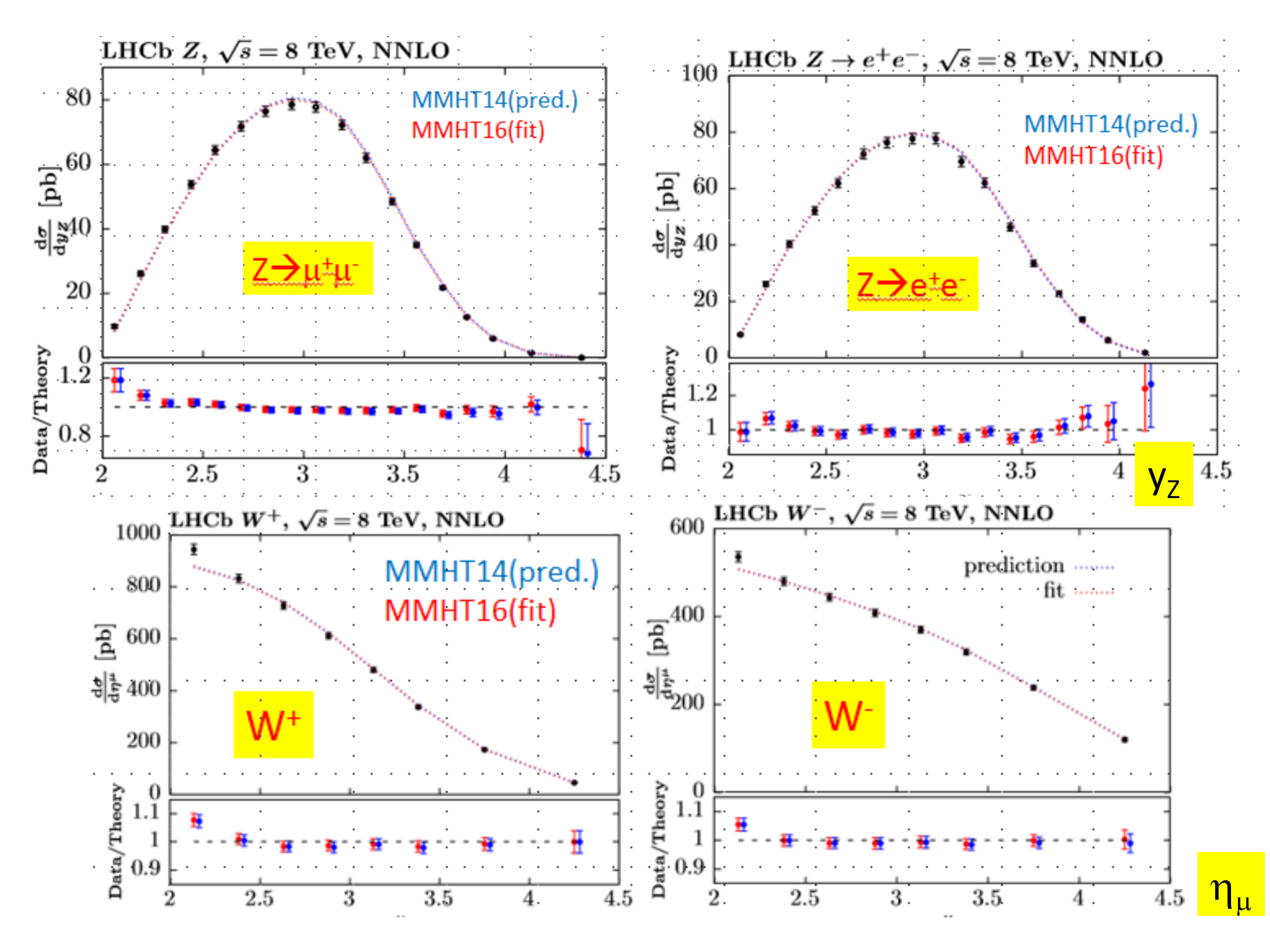} 
 \caption{\sf The NNLO  MMHT14 prediction and the MMHT16 fit to the 8 TeV LHCb data \cite{LHCbWZ8}. }
 \label{fig:LHCbWZ}
\end{center}
\end{figure}
\begin{figure} [H]
 \begin{center}
 \label{fig:CMSW} 
 \includegraphics[clip=true,trim=.0cm 5.cm 0.cm 5.0cm,height=6.0cm]{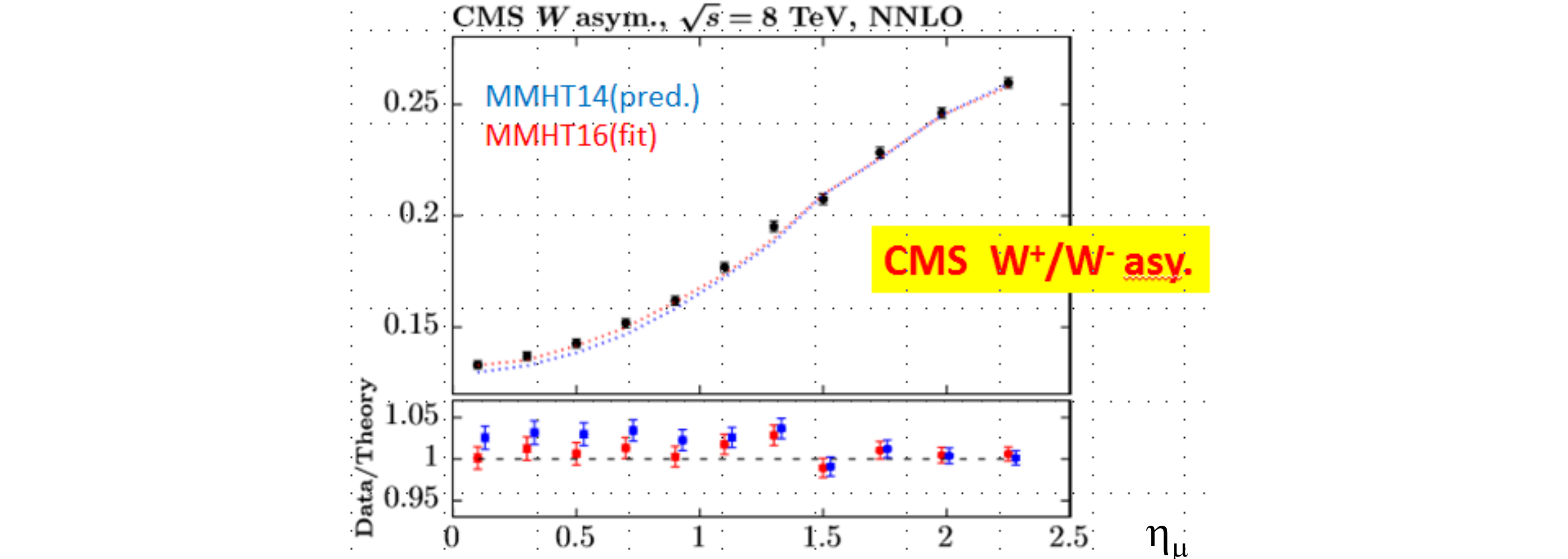}
 \caption{\sf The MMHT14 prediction and the MMHT16 fit to the CMS charge asymmetry data \cite{CMSW}. The fit was to the individual distributions, not the asymmetry. }
\label{fig:CMSW} 
\end{center}
\end{figure}
\begin{figure}[H]
 \begin{center}
 \includegraphics[clip=true,trim=0.0cm 1.0cm 0cm 1.0cm,width=15cm]{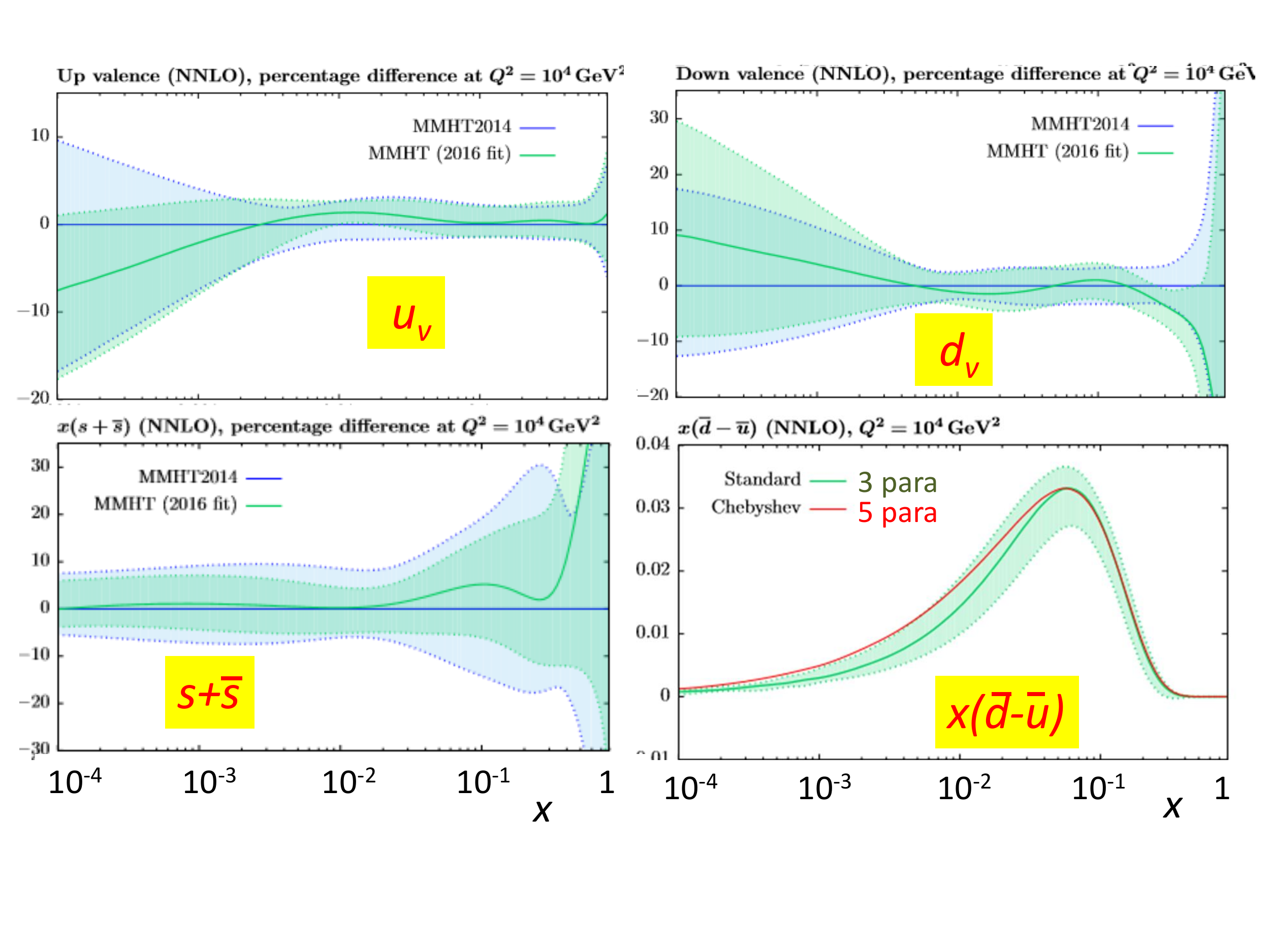} 
 \vspace{-1.0cm}
 \caption{\sf The change in the central values and the reduction of the uncertainty in $u_V$, $d_V$ and $(s+\bar{s})$ at $Q^2=10^4$ GeV$^2$. in going from NNLO MMHT14 to MMHT16 gobal analysis.  The fourth plot is the behaviour of $x(\bar{d}-\bar{u}$) at $Q^2=10^4$ GeV$^2$ in the preliminary NNLO MMHT16 fit -- the red curve is the result of increasing the parameterization of $x(\bar{d}-\bar{u}$) to five parameters }
\label{fig:Fig1MMHTdiff}\end{center}
\end{figure}

The MMHT16 fit shows no tension, so the PDFs are very similar to those of MMHT14. Indeed, $\chi^2$ increases by only 15 for the remainder of the data. 
The only significant change is in $u_V$ and $d_V$, and a reduction in the uncertainty of ($s+\bar{s}$) due to the data of \cite{CMSWc},  see Fig.~\ref{fig:Fig1MMHTdiff}. These changes are to be expected from the inclusion of the more precise $W^\pm$ and $Z$ LHC data in the fit, since they probe the quark distributions in an $x$ region from $2\times 10^{-4}$ to 0.5
 
 Since there was a claim \cite{ABM} that $x(\bar{d}-\bar{u})$ surprisingly preferred a negative value as $x\to 0$, we investigated this further by extending the ($\bar{d}-\bar{u}$) parameterization from 3 to 5 free parameters. The result shows that ($\bar{d}-\bar{u})\to 0$ as $x\to 0$, see the last plot of Fig.~\ref{fig:Fig1MMHTdiff}.  Also we see there is no inclination for ($\bar{d}-\bar{u}$) to go negative in a small region around $x=0.3$, which was a feature of earlier fits. Finally, if the coupling is left free then $\alpha_S(M_Z^2)\simeq 0.118$ as compared to 0.1172 for MMHT14, and we obtain a good description of the $\sigma(t{\bar t})$ data with $m_t^{\rm pole}$=173.4 GeV.

\section{PDFs with QED corrections}

For the level of accuracy that we are now approaching, it is important to account for electroweak corrections. That is, we need PDFs which incorporate QED into the evolution -- in other words, we need to include the photon PDFs, $\gamma_{p,n}(x,Q^2)$ of the proton and neutron. Previous MRST2004QED sets \cite{MRSTqed} assumed that the $\gamma(x,Q^2)$ partons were generated by photon emission off a model for valence quarks with QED evolution from $m_q \to Q_0$. 
The most direct measurement of the photon PDF at that time was wide-angle scattering of the photon by an electron beam via the process $ep \to e\gamma X$, where the final state electron and photon are produced with equal and opposite large transverse momentum.  The MRST2004QED photon PDF was in agreement with the existing ZEUS measurement of this process.
Recent sets published by NNPDF \cite{NNPDF} and CT \cite{CT} have large uncertainties for $\gamma(x,Q^2)$.

\begin{figure} [h]
\begin{center} 
 \includegraphics[clip=true,trim=0.0cm 9.cm 0cm 6.0cm,width=12.0cm]{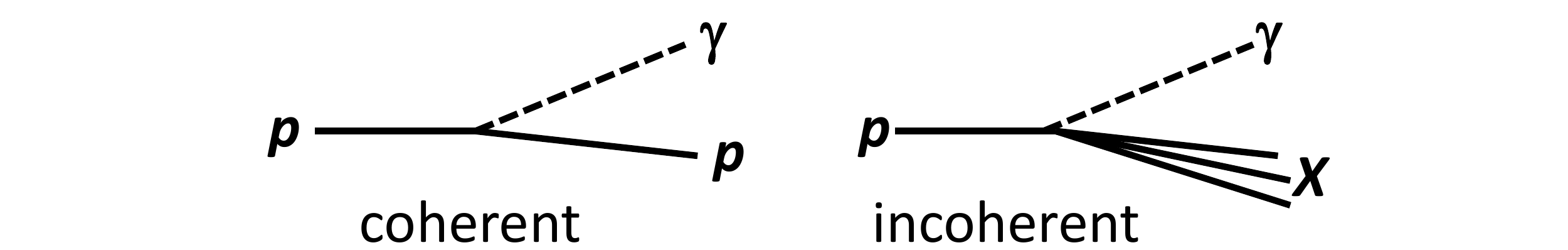} 
 \caption{\sf The coherent and incoherent contributions to the photon PDF, $\gamma_p$, corresponding respectively to photon emission directly from the proton and from an inelastic event.}
\label{fig:photonPDF}
\end{center}
\end{figure}

In a `new' development \cite{MR,HKM,Salam} it was emphasized that the photon PDF, $\gamma_p(x,Q_2)$, is actually quite precisely known.  The distribution is divided into two components $\gamma_p(x,Q_0^2)=\gamma_p^{\rm coh}+\gamma_p^{\rm incoh}$, see Fig.~\ref{fig:photonPDF}.  The first contribution (which comes from coherent photon emission from the `elastic' proton) is accurately known from the form factors of {\it elastic} electron-proton scattering; it is the major part of the input $\gamma_p(x,Q^2_0)$.   Similarly, the incoherent term is constrained by the well-measured structure functions of {\it inelastic} electron-proton scattering for $W^2\gapproxeq 3.5$ GeV$^2$, together with information from the resonance region $W^2\lapproxeq 3.5$ GeV$^2$ \cite{Salam}. Actually this observation is closely related to \cite{HKM}, where it was shown that the incoherent contribution may be determined from DGLAP evolution (including the $\gamma_p$ PDF), which will allow a global parton analysis with a full treatment of uncertainties; this procedure is currently being implemented,
possibly with further improvements suggested by the study in \cite{Salam}.

\bibliographystyle{JHEP}

\bibliography{F}

\end{document}